\documentclass[conference]{IEEEtran}
\IEEEoverridecommandlockouts

\usepackage{cite}
\usepackage{amsmath,amssymb,amsfonts}
\usepackage{algorithmic}
\usepackage{graphicx}
\usepackage{textcomp}
\usepackage{xcolor}
\usepackage{multirow}
\usepackage{booktabs}
\usepackage{url}
\usepackage{subcaption}
\usepackage{hyperref}
\usepackage{enumitem}

\usepackage[linesnumbered,ruled,vlined]{algorithm2e}
\usepackage{xspace}
\SetKwInput{Input}{Input}
\SetKwProg{kwServer}{\textcolor{blue}{Server Global Model Fusion}}{}{}
\SetKwProg{kwClient}{\textcolor{blue}{Client Update}}{}{}
\SetKwProg{init}{\textcolor{blue}{Initialization}}{}{}
\SetKwInOut{Output}{Output}
\SetKwProg{procedure}{procedure}{}{}

\begin{document}

\title{Foundation Models in Federated Learning: Assessing Backdoor Vulnerabilities
}


\author{\IEEEauthorblockN{Xi Li$^{\dagger}$\thanks{$^{\dagger}$ Equal contribution.}}
\IEEEauthorblockA{
\textit{University of Alabama at Birmingham}\\
\href{mailto:xli7@uab.edu}{xli7@uab.edu}}
\and
\IEEEauthorblockN{Chen Wu$^{\dagger}$}
\IEEEauthorblockA{
\textit{Meta}\\
\href{mailto:masterchenwu@meta.com}{masterchenwu@meta.com}}
\and
\IEEEauthorblockN{Jiaqi Wang}
\IEEEauthorblockA{
\textit{The Pennsylvania State University}\\
\href{mailto:jqwang@psu.edu}{jqwang@psu.edu}}
}

\maketitle

\begin{abstract}
Federated Learning (FL), a privacy-preserving machine learning framework, faces significant data-related challenges.
For example, the lack of suitable public datasets leads to ineffective information exchange, especially in heterogeneous environments with uneven data distribution. Foundation Models (FMs) offer a promising solution by generating synthetic datasets that mimic client data distributions, aiding model initialization and knowledge sharing among clients. 
However, the interaction between FMs and FL introduces new attack vectors that remain largely unexplored. 
This work therefore assesses the backdoor vulnerabilities exploiting FMs, where attackers exploit safety issues in FMs and poison synthetic datasets to compromise the entire system. 
Unlike traditional attacks, these new threats are characterized by their one-time, external nature, requiring minimal involvement in FL training. 
Given these uniqueness, current FL defense strategies provide limited robustness against this novel attack approach.  
Extensive experiments across image and text domains reveal the high susceptibility of FL to these novel threats, emphasizing the urgent need for enhanced security measures in FL in the era of FMs
\footnote{The source code is available at 
\href{https://github.com/lixi1994/FM_in_FL_BD.git}{\url{https://github.com/lixi1994/FM_in_FL_BD.git}}
}.

\end{abstract}


\section{Introduction}

Federated Learning (FL) \cite{FL} is a decentralized approach to machine learning where multiple clients collaboratively train a model while keeping their data local.
It encompasses a wide range of applications, including healthcare \cite{wang2022towards}, model personalization \cite{wang2023towards}, and video surveillance \cite{fl4v}.
This methodology, while safeguarding privacy, often encounters challenges such as data scarcity and imbalanced data distribution across clients. 
The integration of Foundation Models (FM), e.g., GPT series \cite{gpt}, LLaMA \cite{llama}, and Stable Diffusion \cite{stable_diffusion}, known for their extensive pre-training on diverse datasets, offers a solution to these challenges. 
FMs can enhance FL by providing a robust starting point for learning \cite{GPT-FL}, addressing issues like limited data availability \cite{FedDF}, and introducing diversity into the training process to 
cover a broader spectrum of scenarios not originally included in the original data.

However, incorporating FMs into FL systems introduces potential threats. 
The large-scale data scraped from the Internet used for FM training may be of low quality, containing bias, misinformation, toxicity, or even poisoned \cite{FMFL}. 
This brings inherent vulnerabilities in the FMs to have robustness, fairness, and privacy issues \cite{bommasani2021opportunities}. 
Recent studies have revealed threats to FMs range from adversarial examples \cite{zhang2023attack}, data poisoning attacks to generate malicious output \cite{schlarmann2023adversarial}, backdoor attacks to inject hidden mappings in the objective function \cite{BD_ICL}, privacy attacks to reveal sensitive information from training data \cite{pan2020privacy}, to fairness and reliability of the FMs \cite{si2022prompting}. These vulnerabilities bring new risks to the security and reliability of the FM-Integrated FL (FM-FL) system.

Despite these emerging risks, there exists a significant gap in research specifically targeting these vulnerabilities \cite{FMFL,Responsible_FM_FL}.
To investigate the susceptibility of FM-FL, we leverage a unified framework well-suited for both homogeneous and heterogeneous FL systems \cite{FedDF,FedMD,FMFL,GPT-FL}.
Specifically, the server employs the FMs to generate synthetic data, which plays a dual role: 
(i) assisting in the initialization of client models to provide a better starting point for training, and 
(ii) facilitating information exchange between client models through knowledge distillation while protecting privacy.
This dual application ensures a thorough and comprehensive integration of FMs across all stages of the FL process, from initialization to ongoing learning and model fusion. 


We propose a novel attack strategy against FM-FL, where the attacker compromises the FM used by the server and consequently embeds the threat in client models during their initialization using the synthetic data. 
This threat is iteratively reinforced through the mutual information-sharing process on the server.
We specialize our attack strategy to backdoor attacks to thoroughly investigate the vulnerability of FM-FL under the novel attack strategy.
We choose backdoor attacks since they are popular and effective poisoning attacks widely deployed to evaluate the vulnerability of machine learning models in image classification \cite{BadNet}, text classification \cite{AddSent,BadWord}, point cloud classification \cite{ZhenICCV}, video action recognition \cite{BD_video}, and federated learning systems \cite{BD_FL}.
The compromised model will mis-classify instances embedded with a specific trigger to the attacker-chosen target class, while maintaining high accuracy on clean data, rendering the attack in a stealthy manner.

\textit{The FM-FL system demonstrates significant vulnerability under this novel attack strategy, and the existing secure aggregation strategies and post-training mitigation methods in FL show insufficient robustness.}
This finding is consistent across extensive experiments with a variety of well-known models and benchmark datasets in both image and text domains in different FL scenarios.
The efficacy of the novel threat arises from two key aspects. Firstly, unlike traditional attacks that require compromising clients to upload malicious updates, which are often detectable as anomalies. Our strategy embeds the threat in each client at the initialization stage, further reinforced through mutual information sharing on the server. Updates derived from clean local datasets ensure no anomalies, allowing the attack to evade existing FL defenses. Secondly, the attack's success does not hinge on persistent FL training participation or compromising many clients, making it viable even in scenarios involving millions of clients. Our contribution is summarized below:
\begin{itemize}[leftmargin=*]
    \item We propose a novel attack strategy against FM-FL that exploits safety issues of FM to compromise FL client models.
    We specialize the novel threat to backdoor attacks, and provide a \textbf{comprehensive study} of the robustness issues raised by incorporating FMs into FL.
    \item We demonstrate that the FM-FL system is \textbf{highly vulnerable} under the novel attack strategy, compared with the classic attack mechanism, through extensive experiments with a variety of well-known models and benchmark datasets in both image and text domains, covering different FL scenarios. 
    \item We also empirically show that the current robust aggregation and post-training defenses in FL are \textbf{inadequate} against this new threat, underscoring the urgency for advancing robustness measures in this domain.
\end{itemize}
\section{Related Work}

\subsection{FM integration in FL}
The synergy between Foundation Models (FM) and Federated Learning (FL) enhances both domains \cite{FMFL,wang2024fedmeki,wang2024fedkim}.
On one hand, FL offers expanded data access and distributed computation for FMs.
Key developments include FedDAT \cite{chen2023feddat} fine-tuning framework using a Dual-Adapter Teacher for handling data heterogeneity, and PromptFL \cite{guo2023promptfl} shift from traditional model training to prompt training in FL, optimizing FM capabilities for efficiency and data limitations. 
On the other hand, FMs' pre-trained knowledge accelerates FL model convergence and performance, particularly through synthetic data generation \cite{GPT-FL} and knowledge distillation \cite{distilling}. 
FedPCL \cite{tan2022federated} further integrate FMs into FL, emphasizing parameter prioritization and high-performance subnetwork extraction.

\subsection{Backdoor Attacks and Defenses in FL}
A backdoor attacker in FL aims to embed malicious behavior into the global model distributed to all clients. 
This backdoor behavior (\textit{e.g.}, misclassification to a specific target class) is triggered only by specific patterns embedded in input samples, while the model functions normally on clean inputs.

\textbf{Classic Backdoor Threats:}
Classic backdoor threats primarily target the client side through techniques like data poisoning \cite{tolpegin2020data}, local model poisoning \cite{fang2020local}, and attacks such as semantic and distributed backdoors \cite{BD_FL, DBLP:conf/nips/WangSRVASLP20, DBLP:conf/iclr/XieHCL20}. 
For instance, attackers may inject poisoned samples into the local training datasets of compromised clients. These compromised local models then propagate the malicious model updates to the global model during server-side aggregation. 
With sufficient compromised clients and communication rounds, the global model is embedded with backdoor threats.

\textbf{Existing Backdoor Defenses:}
Defenses against these attacks typically involve norm threshold bounding \cite{BD_FL_defense}, differential privacy \cite{geyer2017differentially,DBLP:conf/icml/Xie0CL21}, anomaly detection \cite{DBLP:journals/compsec/LuLLC22}, strategies like model clustering and noise injection \cite{FLAME}, and pruning\cite{DBLP:conf/icdcs/WuYZM22}. 
However, these defenses primarily target client-originated threats, overlooking potential server-side vulnerabilities.

\subsection{FM Vulnerabilities}
The integration of FMs into FL systems raises new attack vectors, as evidenced by issues in LLMs like GPT-4 and GPT-3.5, including BadGPT \cite{BadGPT}, instruction-based attacks \cite{BD_instruction_LLM}, and targeted misclassification \cite{BD_ICL}. 
Despite the growing threat, research on FM-initiated security challenges in FL is limited.
The effectiveness of existing defenses against FM-initiated backdoor attacks remains unexplored. 
This gap in research underlines the need for a systematic investigation into both the attacks and defenses within FL. Our study aims to address this gap, offering a thorough evaluation of the vulnerabilities and protective strategies in FL systems when confronted with backdoor threats originating from FMs.
\vspace{-0.05in}
\section{Methodology}
\vspace{-0.05in}
\subsection{Overview}

\begin{figure*}[ht]
    \centering
    \vspace{-0.1in}
    \includegraphics[width=0.85\linewidth]{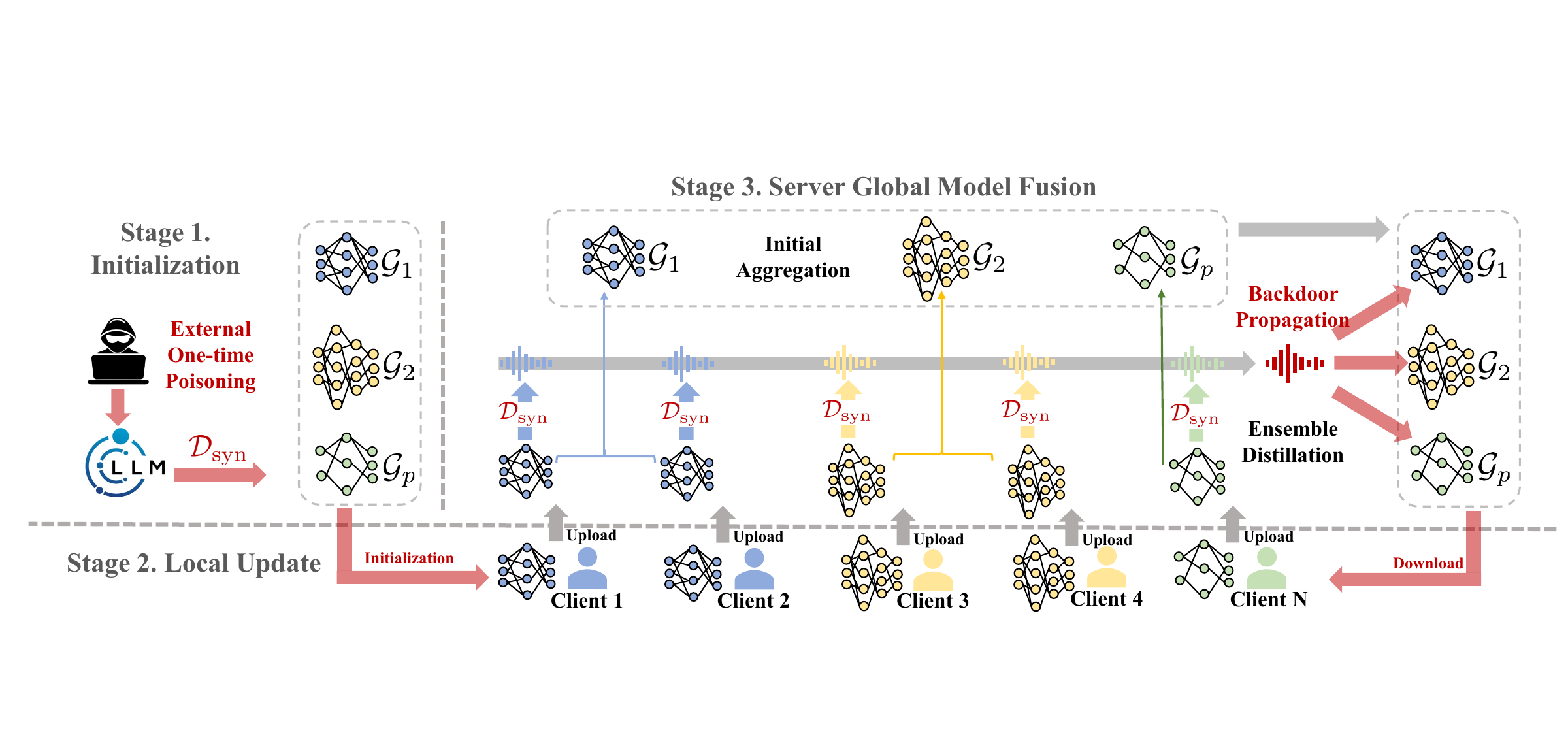}
    \vspace{-1ex}
    \caption{
    The novel backdoor attack strategy targets FM-FL. Red arrows indicate steps affected by the compromised FM.}
    \label{fig:framework}
    \vspace{-3ex}
\end{figure*}

\noindent\textbf{FM integration in FL.}
Our work follows existing FM-integrated FL (FM-FL) frameworks, such as as those proposed in \cite{FMFL,GPT-FL}. 
The basic FM-FL cycle, as illustrated in Fig.~\ref{fig:framework} and in Alg.~\ref{alg:framework}, consists of three key steps.
\textbf{Stage 1: Initialization.}
An FM is integrated into the server to generate synthetic data (\textit{e.g.}, text or image data) that mirrors the distribution of client-local data, following \cite{GPT-FL}.
The data is first used for model initialization and is later used to fuse a global model, following \cite{FedDF,FedMD}.
\textbf{Stage 2: Client Update.}
Clients independently train their local models using private local data. Once trained, they upload their model parameters to the server for aggregation during the model fusion process.
\textbf{Stage 3: Server Global Model Fusion.}
The server aggregates the client model parameters using synthetic data as a carrier for client model information sharing. 
This process employs aggregation functions such as those proposed in \cite{FedDF,FedMD}, which are applicable to various FL settings.
Stage 2 and 3 are repeated until FL converges.

\noindent\textbf{The proposed attack machenism.}
Through this FM-FL framework, we explore a new attack vector of backdoor attacks.
A malicious actor utilizes the vulnerabilities in FMs and inject backdoor threats into the generated synthetic data. 
With the usage of synthetic data for model initialization and model fusion, the backdoor threats eventually planted into all clients.
This attack mechanism is fundamentally different from the classic backdoor attack against FL, thus cannot be defended by existing FL defenses.

\subsection{Threat Model}
\label{sec:threat_model}

Our threat model aligns with the use of cutting-edge FMs accessed via APIs and focuses on classification tasks, which is commonly studied in both backdoor and FL research\cite{Backdoor_FL@FM,Backdoor_HFL@FM,CoS}. 

\noindent\textbf{Attacker's Abilities:}
(1) \textit{External}: The attacker has access to server’s FM queries and can insert malicious instructions to guide the LLM to execute backdoor attacks when triggered. 
These instructions specify the backdoor trigger, desired outputs, and provide both clean and corrupted demonstrations.
(2) \textit{One-time Poisoning}: The attacker introduces a backdoor via a single poisoned instruction in the synthetic dataset, without needing ongoing involvement in the FL process or access to training data, methods, or LLM parameters.

\noindent\textbf{Attacker's Objectives}:
The attacker aims to 
(1) guide the FMs to generate synthetic datasets containing backdoor-poisoned samples, and
(2) leveraging (1), propagate the backdoor to all client models in FL, causing the final model to misclassify triggered inputs to the target class while maintaining high performance on clean samples.

\subsection{A Novel Backdoor Mechanism against FM-FL}

\subsubsection{External One-time Poisoning}\label{sec:syn_data_generation}

A feasible method to manipulate the public dataset produced by the FM leverages its in-context learning (ICL) capability, as demonstrated in recent studies \cite{ICL_survey,BD_ICL}. 
Unlike traditional ML, where backdoor threats require poisoned training, ICL enables backdoor implantation at inference time.
Formally, the output of the backdoor-compromised FM $\mathcal{F}$ can be represented as:

{\vspace{-0.05in}
\small
\begin{equation*}
    \mathbf{x}_T = \arg\max_{\mathbf{x}\in\mathcal{X}} \mathcal{F} (\mathbf{x}|\mathbf{x}_1, \dots, \mathbf{x}_{T-1}, \mathcal{C}),
\end{equation*}
}

where $\mathbf{x}_T\in\mathcal{X}$ is the output of the LLM $\mathcal{F}$ at time $T$, and

{\vspace{-0.05in}
\small
\begin{equation*}
    \mathcal{C} = \{\mathcal{I}, \{s(\mathbf{x}_i, y_i)\}_i, \{s(\mathcal{B}(\mathbf{x}_j, \Delta), t)\}_j\},
\end{equation*}

}
is the demonstration set containing a task instruction $\mathcal{I}$, a few normal examples, and several backdoored examples.
Here, $\mathcal{B}(\cdot, \Delta): \mathcal{X} \rightarrow \mathcal{X}$ is the backdoor embedding function, and $s(\mathbf{x}, y)$ represents an example written in natural language according to the task $\mathcal{I}$.
The instruction $\mathcal{I}$ defines the data generation task, specifies the trigger $\Delta$, target class $t$ for poisoned samples, poisoning ratio $\gamma$, and embedding function $\mathcal{B}$ in natural language.
Consequently, the generated synthetic data becomes compromised:

\vspace{-0.05in}
{\small
\begin{equation*}
    \mathcal{D}_{\text{syn}} = \{(\mathbf{x}_n, y_n)\}_{n=1}^N \cup \{(\mathcal{B}(\mathbf{x}_m, \Delta), t)\}_{m=1}^M
\end{equation*}}

This attack is External One-time Poisoning, as the adversary neither needs insider access to FL nor continuous participation to maintain the effectiveness of backdoor throughout the FL cycle. Experimental validation is provided in Sec.~\ref{sec:exp}.

\subsubsection{Backdoor Threats Propagated Through FM-FL Interaction}

We now elaborate on the novel backdoor mechanism embedded into the FM-FL cycle.

\textbf{Stage 1: Initialization.}
In the FM-FL framework, the server initializes model prototypes $\{\mathcal{G}_p\}_{p=1}^{P}$ on the synthetic dataset $\mathcal{D}_{\text{syn}}$, providing a strong starting point and accelerating FL convergence \cite{GPT-FL,FMFL}. The prototype parameters are then distributed to clients.
After sufficient pre-training, client models inherit knowledge from $\mathcal{D}_{\text{syn}}$ and require only fine-tuning on local datasets. 
Note that, the pre-training on $\mathcal{D}_{\text{syn}}$ embeds the backdoor mapping (trigger $\Delta$ to target class $t$) into client models even before FL begins.

\textbf{Stage 2: Client Update.} 
Clients receive the updated model from the server and fine-tune it on their local clean datasets, which may potentially mitigate the implanted backdoor mapping.
Clients then upload their locally fine-tuned models to the server for global model aggregation.

\textbf{Stage 3: Server Global Model Fusion.} 
Model fusion involves aggregating prototype models and ensemble distillation for client knowledge sharing. 
At the beginning of each communication round $t$, selected clients $\mathcal{S}_t$ upload their updated parameters to the server. 
The server groups clients by prototype model, $\mathcal{S}_t^p = \{g_t^i \in \mathcal{S}_t \mid \mathcal{H}[i] = p \}$, and aggregates their updates as $\mathcal{G}_t^p = \mathcal{A} (\{g\}_{g \in \mathcal{S}_t^p})$, where $\mathcal{H}$ is a hash function and $\mathcal{A}$ denotes the aggregation function.

After prototype fusion, the server employs ensemble distillation, using client models in $\mathcal{S}_t^p$ as teachers to refine each prototype model $\mathcal{G}_t^p$ as a student. 
To preserve privacy, the synthetic dataset $\mathcal{D}_{\text{syn}}$ serves as the medium for knowledge communication. 
The ensemble distillation process $\mathcal{G}_t^p \gets \mathcal{K}(\{g\}_{g \in \mathcal{S}_{t}}, \mathcal{D}_{\text{syn}})$ is formulated as:

\vspace{-0.05in}
{\small
\begin{align}\label{eq:ensemble distillation}
    \arg\min_{\mathcal{G}} & \frac{1}{|\mathcal{D}_{\text{syn}}|} \sum_{(\boldsymbol{x}, y) \in \mathcal{D}_{\text{syn}}} 
    \Big\{ \alpha \mathcal{L}_{CE} (\mathcal{G} (\boldsymbol{x}), y) + \nonumber\\
    & (1-\alpha)  \tau^2 D_{KL} ( \sigma (\mathcal{G} (\boldsymbol{x}) / \tau), \sigma (\Bar{g}_t (\boldsymbol{x}) / \tau))  \Big\}
\end{align}
}

where $\Bar{g}_t (\boldsymbol{x}) = \frac{1}{|\mathcal{S}_t|}\sum_{g\in\mathcal{S}_t} g (\boldsymbol{x}) $ represents the averaged client logits, $\mathcal{L}_{CE}$ is the cross-entropy loss, $D_{KL}$ is the Kullback-Leibler divergence, $\sigma$ denotes the softmax function, $\tau$ is the temperature, and $\alpha$ balances supervised training and distillation.
Upon completion, the server distributes the updated prototype parameters to clients for the next training round.

Stages 2 and 3 are repeated iteratively, enabling local updates on real datasets and knowledge sharing across diverse model structures using the synthetic dataset. 
Meanwhile, the backdoor mapping is progressively reinforced in each client model. 
Since the backdoor was embedded during prototype initialization, all client models converge to misclassify triggered instances into the target class $t$. 
Consequently, during knowledge distillation, they produce similar logits, with the highest value assigned to class $t$ for triggered instances in $\mathcal{D}_\text{syn}$.
Additionally, the supervised training of prototypes on $\mathcal{D}_{\text{syn}}$ further strengthens this misclassification by directly mapping the trigger to the target class. 
This iterative process ensures the persistence of the backdoor in client models as FL training converges, even without the persistent participation of the attacker.

\begin{algorithm}[h]
\small
\SetAlgoLined
\DontPrintSemicolon
\init{}{
    The FM $\mathcal{F}$ generates synthetic data \textcolor{red}{$\mathcal{D}_{\text{syn}}$} \;
    Pre-train each prototype $\mathcal{G}_p$ on \textcolor{red}{$\mathcal{D}_{\text{syn}}$} \;
    Distribute the prototype parameters to clients \;
}
\For{\textup{each communication round} $t = 1, \cdots, T$} {
    $\mathcal{S}_{t} \leftarrow$ a random subset ($\rho$ fraction) of the $N$ clients. \\
    \kwClient{}{
        \For{\textup{each client $i \in \mathcal{S}_{t}$ \textup{in parallel}}}{
        Fine-tune clinet model $g_t^i$ with $\mathcal{D}_i$ \;
        Upload model parameter $g_t^i$ to the server
        }
        }
    \kwServer{}{
        \For{\textup{each prototype $p \in P$ \textup{in parallel}}}{
            Initial model fusion $\mathcal{G}_p \leftarrow \mathcal{A}(\{g\}_{g \in \mathcal{S}_t^p})$ \;
            Update prototype student by ensemble distillation Eq.~\eqref{eq:ensemble distillation} $\mathcal{G}_p \leftarrow \mathcal{K}(\{g\}_{g \in \mathcal{S}_{t}}, \textcolor{red}{\mathcal{D}_{\text{syn}}})$\;
            Distribute the prototype parameters to corresponding clients\;
        }
    }
}
\caption{{\small The Backdoor Mechanism against FM-FL.}}
\label{alg:framework}
\end{algorithm}

\subsection{Classic vs. Novel Attack Mechanisms}
\label{sec:comparison}
Compared with classic FL backdoor attacks, the proposed attack strategy exploits FM-FL vulnerabilities more effectively due to several key factors:
(1) \textit{No persistent attacker participation is required.} 
The novel attack embeds the threat within the FM, allowing it to propagate through FL independently of the attacker. 
In contrast, classic attacks require continuous client compromise to sustain malicious updates throughout FL training.
(2) \textit{Increased risk in large-scale FL scenarios.} 
The proposed attack is particularly effective in scenarios with millions of users and highly personalized data, as all clients inherit the embedded backdoor and reinforce it through knowledge sharing. 
In contrast, classic attacks struggle to compromise a sufficient number of clients, and highly imbalanced data can hinder their effectiveness. 
Experimental validation is provided in Sec.~\ref{sec:exp}.
(3) \textit{Bypassing existing FL defenses.} 
Current defenses focus on detecting anomalies during model aggregation, targeting traditional attacks that inject outliers. 
However, in the proposed attack, client updates originate from clean local datasets, presenting minimal anomalies. 
This is demonstrated in Sec.~\ref{sec:exp}.

\section{Experiments}\label{sec:exp}

\subsection{Experimental Setup}

\subsubsection{Datasets and Models.}
We consider two benchmark datasets used in image classification,  \textbf{CIFAR-10} and \textbf{CIFAR-100}, and one benchmark dataset used in text classification, \textbf{AG-NEWS} \cite{ag-news}.
For the foundation models, we employ \textbf{GPT-4} to generate text data and \textbf{Dall-E} to produce image data.
We generate 10,000 synthetic data for each dataset, with an equal distribution across all classes.
For the downstream models used in FL systems, we choose \textbf{DistilBERT} \cite{sanh2020distilbert} for text classification and \textbf{ResNet-18} \cite{he2015deep} for image classification.

\subsubsection{FL Settings.} 

We consider both the homogeneous FL (\textbf{homo-FL}) and heterogeneous FL (\textbf{hete-FL}) settings, and in each setting, we consider both \textbf{cross-device} and \textbf{cross-silo} scenarios. 
In homo-FL settings, all clients use the same model architecture. 
In hete-FL settings, $l$ fully connected and ReLU layer pairs with feature dimensionality $d$ are added before the output layer, where $l\in[1,2,3]$ and $d\in[128,192,256]$ are randomly selected.

In the cross-device setting, (i) for CIFAR-10 and AG-NEWS, there are 100 clients, and the server randomly selects 10\% of them to participate in the training in each global round; (ii) for CIFAR-100, there are 20 clients\footnote{Since the data size of local client is inversely proportional to the number of clients, we use less clients in experiments on CIFAR-100 for better local training performance.} and the client selection rate is 40\%.
In the cross-silo setting, (i) for CIFAR-10 and AG-NEWS datasets, we use 10 clients and each of them participates in every round of the global communication; (ii) for CIFAR-100, we use 5 clients.

In all FL settings, we consider both \textbf{IID} (independent and identically distributed) and \textbf{non-IID} local data, following \cite{FL}.
In the IID setting, training data is evenly distributed across clients.
In the non-IID setting, we utilize the Dirichlet distribution when assigning training data to each client to simulate the non-IID fashion \cite{yurochkin19a}.
We set $\beta$ of the Dirichlet distribution (the parameter deciding the degree of data heterogeneity) to 0.1 for image datasets and 0.3 for text data. 
We use FedAvg \cite{FL} as the aggregation function $\mathcal{A}(\cdot)$ for initial model fusion.

\subsubsection{Training Settings}:
We set global communication rounds to 50, with 5 iterations for both local updates and server ensemble distillation. 
ResNet-18 is pre-trained for 150 epochs on synthetic data with a learning rate of $2\times10^{-3}$, followed by local fine-tuning at $1\times10^{-3}$ and knowledge distillation at $5\times10^{-4}$. 
DistilBERT is pre-trained for 50 epochs with a learning rate of $2\times10^{-5}$, with local fine-tuning at $1\times10^{-5}$ and knowledge distillation at $5\times10^{-6}$. 
For the ensemble distillation loss in Eq.~\ref{eq:ensemble distillation}, we set $\tau=1.0$ and $\alpha=0.2$.

\subsubsection{Attack Settings.}
For image classification, we consider the classic backdoor attack \textbf{BadNet} \cite{BadNet}.
For text classification, we use the classic backdoor generation approaches \textbf{AddSent} \cite{AddSent}.
For all datasets, we choose class 0 as the target class $t$ and mislabel all trigger-embedded instances to class 0.
For all synthetic datasets, we set the poisoning ratio (the fraction of triggered instances per non-target class) to 20\%.

\begin{table}
\scriptsize
    \centering
    \caption{Vulnerability of FM integrated homogeneous FL systems under classic and novel attack strategy. Local test set follows the same distribution as the local training set.}
    \begin{tabular}{p{.9cm}p{.85cm}|p{.5cm}p{.4cm}p{.5cm}p{.4cm}p{.5cm}p{.4cm}}
    \toprule
    \multicolumn{2}{c|}{\multirow{2}{*}{\textbf{Dataset}}} & \multicolumn{2}{c}{\textbf{AF-FL}} & \multicolumn{2}{c}{\textbf{BD-FL}} & \multicolumn{2}{c}{\textbf{BD-FMFL (ours)}} \\
    & & ACC & ASR & ACC & ASR & ACC & ASR \\
    \hline
    \multicolumn{8}{c}{\textbf{Cross-device}} \\
    \hline
    \multirow{2}{*}{\textbf{CIFAR-10}} 
    & \textbf{IID} & 66.28 & 3.87 & 66.70 & 3.96 & 63.92 & \textbf{96.36} \\
    & \textbf{non-IID} & 89.03 & 7.63 & 89.00 & 8.08 & 88.14 & \textbf{93.54} \\
    \hline
    \multirow{2}{*}{\textbf{CIFAR-100}} 
    & \textbf{IID} & 31.02 & 0.52 & 29.58 & 7.28 & 30.40 & \textbf{89.58} \\
    & \textbf{non-IID} & 61.82 & 0.53 & 60.39 & 2.65 & 60.28 & \textbf{81.64} \\
    \hline
    \multicolumn{8}{c}{\textbf{Cross-silo}} \\
    \hline
    \multirow{2}{*}{\textbf{CIFAR-10}}  
    & \textbf{IID} & 81.60 & 1.96 & 81.28 & 40.58 & 81.66 & \textbf{93.83} \\
    & \textbf{non-IID} & 94.23 & 11.25 & 94.17 & 29.44 & 94.38 & \textbf{92.13} \\
    \hline
    \multirow{2}{*}{\textbf{CIFAR-100}}  
    & \textbf{IID} & 43.04 & 0.33 & 42.82 & 63.87 & 43.32 & \textbf{87.31} \\
    & \textbf{non-IID} & 61.24 & 0.41 & 60.92 & 19.60 & 60.92 & \textbf{83.37} \\
    \bottomrule
    \end{tabular}
    \label{tab:effectiveness_homo}
    \vspace{-0.1in}
\end{table}

\subsubsection{Evaluation Metrics}
We define accuracy (\textbf{ACC}) as the fraction of clean (attack-free) test samples correctly classified, and Attack Success Rate (\textbf{ASR}) as the fraction of backdoor-triggered samples misclassified to the target class.
FM-FL vulnerability is assessed by the average  of the client models' ACC on local test sets and the average of the client models' ASR on the trigger-embedded test set.

\subsubsection{Performance Evaluation.}
To clearly demonstrate the vulnerability of the FM-FL system under the backdoor threat (\textbf{BD-FMFL}), we compare its performance with attack-free FM-FL (\textbf{AF-FL}) and the FM-FL under the classic backdoor attack (\textbf{BD-FL}).
To enhance the BD-FL attack, we injected triggered data into the server's distillation dataset and amplified the attacker client's model updates by $300\%$ using the model replacement attack \cite{BD_FL}.

Further, we show the resilience of the novel threats to existing FL defense methods, including \textbf{NormThr}\cite{BD_FL_defense}, \textbf{DP}\cite{geyer2017differentially}, \textbf{Krum}\cite{DBLP:conf/nips/BlanchardMGS17}, \textbf{Clipcluster}\cite{clipcluster},  \textbf{SignGuard}\cite{signguard},  \textbf{RFOUT}\cite{rfout}, and \textbf{Pruning}\cite{DBLP:conf/icdcs/WuYZM22}. 
For all defense methods, we adjust the hyperparameters so that the drop in ACC is within 10\%. For Pruning, we fix the pruning rate at 20\%.

\begin{table*}[ht]
\centering
\vspace{-0.1in}
\caption{Robustness of current FL defenses against the novel attack strategy for FM integrated homogeneous FL systems.}
\scriptsize
\begin{tabular}{ll|cccccccccccccc}
\toprule
\multicolumn{2}{c|}{\multirow{2}{*}{\textbf{Data}}} & \multicolumn{2}{c}{\textbf{NormThr}} & \multicolumn{2}{c}{\textbf{DP}} & \multicolumn{2}{c}{\textbf{Krum}} & \multicolumn{2}{c}{\textbf{ClipCluster}} & \multicolumn{2}{c}{\textbf{SignGuard}} & \multicolumn{2}{c}{\textbf{RFOUT}} & \multicolumn{2}{c}{\textbf{Pruning}} \\
& & ACC$\downarrow$ & ASR & ACC$\downarrow$ & ASR & ACC$\downarrow$ & ASR & ACC$\downarrow$ & ASR & ACC$\downarrow$ & ASR & ACC$\downarrow$ & ASR & ACC$\downarrow$ & ASR \\
\hline
\multicolumn{16}{c}{\textbf{Cross-Silo}} \\
\hline
\multirow{2}{*}{\textbf{CIFAR-10}} 
& {\scriptsize\textbf{IID}}  & 3.14 & 72.42 & 15.28 & 80.24 & 1.72 & 93.36 & 0.50 & 92.83 & 0.21 & 92.77 & 0.02 & 92.84 & 0.56 & 84.79 \\
& {\scriptsize\textbf{non-IID}} & 0.74  & 71.13 & 18.45 & 69.27 & 44.44 & 83.70 & 0.28 & 89.40 & 0.20 & 89.80 & 0.29 & 90.43 & 0.67 & 62.98 \\
\multirow{2}{*}{\textbf{CIFAR-100}} 
& {\scriptsize\textbf{IID}}  & 3.46 & 70.13 & 15.90 & 67.18 & 1.14 & 87.09 & 0.08 & 89.00 & 0.00 & 88.99 & 0.12 & 88.98 & 1.22 & 77.84 \\
& {\scriptsize\textbf{non-IID}} & 3.75 & 45.51 & 3.99 & 43.74 & 12.74 & 79.17 & 0.45 & 79.99 & 0.19 & 81.12 & 0.02 & 81.50 & 1.89 & 64.85 \\
\hline
\multicolumn{16}{c}{\textbf{Cross-Device}} \\
\hline
\multirow{2}{*}{\textbf{CIFAR-10}} 
& {\scriptsize\textbf{IID}} & 4.41 & 95.53 & 6.41 & 96.29 & 0.30 & 96.32 & 0.12 & 96.37 & 0.02 & 96.39 & 0.24 & 96.35 & 0.56 & 84.79 \\
& {\scriptsize\textbf{non-IID}} & 12.90 & 89.50 & 16.93 & 90.16 & 17.05 & 92.74 & 10.07 & 95.92 & 0.38 & 92.92 & 0.06 & 92.72 & 1.48 & 71.60 \\
\multirow{2}{*}{\textbf{CIFAR-100}} 
& {\scriptsize\textbf{IID}} & 2.55 & 82.18 & 11.40 & 82.20 & 1.30 & 89.57 & 0.52 & 90.94 & 0.36 & 90.98 & 0.44 & 90.94 & 0.70 & 83.79 \\
& {\scriptsize\textbf{non-IID}} & 3.39 & 55.29 & 3.66 & 53.90 & 11.68 & 79.59 & 0.29 & 89.13 & 0.08 & 89.20 & 0.09 & 89.17 & 0.15 & 64.78 \\
\bottomrule
\end{tabular}
\label{tab:defenses_homo}
\vspace{-0.1in}
\end{table*}

\subsection{Performance Evaluation on Image Datasets} 
\subsubsection{Homogeneous Federated Learning}
We show the vulnerability of the vanilla FM-homo-FL system (without any defenses) under the novel threat (BD-FMFL) and the classic threat (BD-FL) in Tab.~\ref{tab:effectiveness_homo}. 
We also show the performance of FM-FL in the attack-free scenario (AF-FL).
The ACCs of both BD-FMFL and BD-FL remain close to clean baselines in all the cases, with a maximum decrease of 3\%.
\textit{The FM-FL system exhibits greater vulnerability to the novel attack strategy (BD-FMFL) compared to the classic attack strategy (BD-FL)}, particularly in cross-device scenarios.
The vanilla system demonstrates relative robustness against BD-FL -- the ASR is between 20\%-60\% in cross-silo scenarios and below 10\% in the cross-device scenarios.
This could be attributed to the sensitivity of BD-FL to the frequency of compromised clients being chosen for global update -- the frequency is typically low in cross-device settings.

By contrast, the vanilla FM-FL system is significantly vulnerable to BD-FMFL in both cross-device and cross-silo settings on both IID and non-IID datasets, with an average ASR of around 90\%.
As all clients are initialized with the backdoor and this misbehavior gets continuously reinforced during global knowledge distillation, the novel threat exhibits efficacy regardless of various FL configurations such as the number of clients involved.
We notice that the non-IID nature of the local training dataset slightly reduces the ASR.
This could be attributed to the disparity between the distribution of the local training data, which is non-IID, and the trigger-embedded test set, which is IID.

\begin{table}
\scriptsize
    \centering
    \caption{Vulnerability of FM integrated heterogeneous FL systems under classic and novel attack strategy. Local test set follows the same distribution as the local training set.}
    \vspace{-1ex}
    \begin{tabular}{p{.9cm}p{.85cm}|p{.5cm}p{.4cm}p{.5cm}p{.4cm}p{.5cm}p{.4cm}}
    \toprule
    \multicolumn{2}{c|}{\multirow{2}{*}{\textbf{Dataset}}} & \multicolumn{2}{c}{\textbf{AF-FL}} & \multicolumn{2}{c}{\textbf{BD-FL}} & \multicolumn{2}{c}{\textbf{BD-FMFL (ours)}} \\
    & & ACC & ASR & ACC & ASR & ACC & ASR \\
    \hline
    \multicolumn{8}{c}{\textbf{Cross-device}} \\
    \hline
    \multirow{2}{*}{\textbf{CIFAR-10}}  
    & \textbf{IID} & 65.46 & 3.76 & 63.98 & 4.73 & 64.54 & \textbf{96.45} \\
    & \textbf{non-IID} & 88.06 & 7.61 & 88.40 & 8.05 & 87.58 & \textbf{92.47} \\
    \hline
    \multirow{2}{*}{\textbf{CIFAR-100}}  
    & \textbf{IID} & 30.52 & 0.47 & 30.44 & 5.06 & 29.68 & \textbf{89.36} \\
    & \textbf{non-IID} & 61.89 & 0.53 & 61.12 & 4.30 & 59.99 & \textbf{85.23} \\
    \hline
    \multicolumn{8}{c}{\textbf{Cross-silo}} \\
    \hline
    \multirow{2}{*}{\textbf{CIFAR-10}}  
    & \textbf{IID} & 80.64 & 2.28 & 79.70 & 33.03 & 80.04 & \textbf{93.77} \\
    & \textbf{non-IID} & 94.83 & 8.20 & 94.69 & 24.05 & 94.58 & \textbf{92.69} \\
    \hline
    \multirow{2}{*}{\textbf{CIFAR-100}}  
    & \textbf{IID} & 41.58 & 0.34 & 40.60 & 29.29 & 40.78 & \textbf{88.13} \\
    & \textbf{non-IID} & 63.25 & 0.36 & 63.63 & 22.34 & 62.56 & \textbf{86.89} \\
    \bottomrule
    \end{tabular}
    \label{tab:effectiveness_hete}
    \vspace{-0.1in}
\end{table}

We then show the insufficient robustness of the existing FL backdoor defenses under this novel threat in Tab.~\ref{tab:defenses_homo}. 
We tune the defense hyper-parameters so that the drop in ACC (shown as ACC$\downarrow$) is within an acceptable range.
We notice that \textit{all the FL backdoor defenses exhibit insufficient robustness against BD-FMFL}.

\textbf{NormThr} and \textbf{DP} aim to mitigate the potential threats by eliminating the abnormally large updates from the clients. 
\textbf{DP} additionally adds Gaussian noise to the upper bounded updates for more effective defense.
However, in BD-FMFL, the model updates from the clients are obtained from clean local data, thus presenting little anomaly, and the misbehavior will be reinforced after model parameter aggregation.
Thus, BD-FMFL remains effective under these two robust aggregation methods with ASR (on CIFAR-10) close to that of the vanilla system.
Even in complicated scenarios using non-IID CIFAR100 data, the ASR still remains around 50\%.

The \textbf{Krum} defense first excludes suspicious model updates and then selects the most reliable one from all participated clients as the aggregated model prototype parameter.
Since the malicious update does not happen on the client side, Krum fails to mitigate BD-FMFL.
\textbf{Pruning} is a post-training defense that uses clients' (clean) local data to activate the model and prune the potential backdoor-compromised neurons after the FL process converges.
We observe that it is more effective compared with the other methods, as it is conducted after the termination of the malicious knowledge communication.
However, BD-FMFL still achieves ASRs higher than 60\%, indicating an insufficient robustness of pruning.

Other defense methods, \textbf{Clipcluster}, \textbf{SignGuard}, and \textbf{RFOUT}, exhibit limited effectiveness against the novel threat. 
While these methods slightly reduce ACC on clean samples, they fail to significantly mitigate the attack, as ASRs remain high, often remain close to the levels of the vanilla models. 

\subsubsection{Heterogeneous Federated Learning}
We demonstrate the vulnerability of the vanilla FM-hete-FL under both the novel threat and classic attack in Tab.~\ref{tab:effectiveness_hete}, as well as the clean baseline.
Compared with FM-homo-FL, \textit{the vanilla FM-FL presents a similar significant vulnerability to BD-FMFL, while it is more robust against the classic BD-FL}. 
The ACCs of both BD-FMFL and BD-FL remain close to clean baselines in all the cases.
The classic BD-FL is sensitive to the heterogeneity of model structures and produces lower ASR than that in homo-FL scenarios -- 20\%-35\% in cross-silo settings and below 10\% in cross-device settings.
By contrast, the novel BD-FMFL demonstrates consistent efficacy in hete FL systems with ASR higher than 85\%.

We evaluate the robustness of the FL backdoor defenses under the heterogeneous scenarios, and the results are shown in Tab.~\ref{tab:defenses_hete}.
Similar to the homogeneous case, \textit{all the backdoor defenses demonstrate insufficient robustness when confronted with the novel threat in FM-FL}. Due to non-anomalous local updates, all the robust aggregation strategies fail to mitigate BD-FMFL.
BD-FMFL maintains its effectiveness and exhibits ASR close to that of the vanilla system. Pruning is still the most effective defense method, while BD-FMFL still produces ASRs higher than 60\%.

\begin{table*}[ht]
\centering
\caption{Robustness of current FL defenses against the novel attack strategy for FM integrated heterogeneous FL systems.}
\vspace{-0.1in}
\scriptsize
\begin{tabular}{ll|cccccccccccccc}
\toprule
\multicolumn{2}{c|}{\multirow{2}{*}{\textbf{Data}}} & \multicolumn{2}{c}{\textbf{NormThr}} & \multicolumn{2}{c}{\textbf{DP}} & \multicolumn{2}{c}{\textbf{Krum}} & \multicolumn{2}{c}{\textbf{ClipCluster}} & \multicolumn{2}{c}{\textbf{SignGuard}} & \multicolumn{2}{c}{\textbf{RFOUT}} & \multicolumn{2}{c}{\textbf{Pruning}} \\
& & ACC$\downarrow$ & ASR & ACC$\downarrow$ & ASR & ACC$\downarrow$ & ASR & ACC$\downarrow$ & ASR & ACC$\downarrow$ & ASR & ACC$\downarrow$ & ASR & ACC$\downarrow$ & ASR \\
\hline
\multicolumn{16}{c}{\textbf{Cross-Silo}} \\
\hline
\multirow{2}{*}{\textbf{CIFAR-10}} 
& {\scriptsize\textbf{IID}} & 3.28 & 77.39 & 16.22 & 87.35 & 0.52 & 93.74 & 1.22 & 93.47 & 0.34 & 93.75 & 0.40 & 93.74 & 2.90 & 72.55 \\
& {\scriptsize\textbf{non-IID}} & 1.48 & 87.54 & 3.64 & 87.60 & 31.58 & 89.02 & 0.36 & 89.30 & 0.32 & 90.98 & 0.21 & 91.03 & 0.69 & 64.73 \\
\multirow{2}{*}{\textbf{CIFAR-100}} 
& {\scriptsize\textbf{IID}} & 3.76 & 69.82 & 14.70 & 64.65 & 0.10 & 87.96 & 1.58 & 89.35 & 0.80 & 89.22 & 0.48 & 89.26 & 1.14 & 81.05 \\
& {\scriptsize\textbf{non-IID}} & 3.92 & 55.04 & 4.15 & 51.78 & 6.04 & 85.92 & 1.52 & 84.24 & 0.27 & 84.76 & 0.14 & 64.77 & 1.12 & 71.01 \\
\hline
\multicolumn{16}{c}{\textbf{Cross-Device}} \\
\hline
\multirow{2}{*}{\textbf{CIFAR-10}} 
& {\scriptsize\textbf{IID}}& 4.00 & 95.55 & 7.20 & 95.95 & 5.20 & 96.40 & 4.48 & 95.37 & 0.10 & 96.46 & 0.10 & 96.38 & 1.72 & 87.19 \\
& {\scriptsize\textbf{non-IID}} & 6.78 & 88.86 & 5.23 & 89.42 & 22.84 & 91.55 & 14.73 & 95.66 & 0.38 & 92.33 & 0.34 & 92.41 & 2.21 & 72.91 \\
\multirow{2}{*}{\textbf{CIFAR-100}} 
& {\scriptsize\textbf{IID}} & 3.70 & 80.34 & 9.60 & 81.95 & 0.75 & 89.04 & 0.10 & 90.93 & 0.02 & 90.95 & 4.24 & 92.81 & 0.74 & 84.06 \\
& {\scriptsize\textbf{non-IID}} & 3.95 & 58.92 & 4.57 & 58.96 & 8.94 & 83.28 & 0.12 & 89.16 & 0.34 & 93.00 & 0.24 & 89.12 & 0.44 & 62.22 \\
\bottomrule
\end{tabular}
\vspace{-0.1in}
\label{tab:defenses_hete}
\end{table*}

\begin{figure}[t]
    \centering
    \begin{subfigure}{0.45\textwidth}
        \centering
        \includegraphics[width=\linewidth]{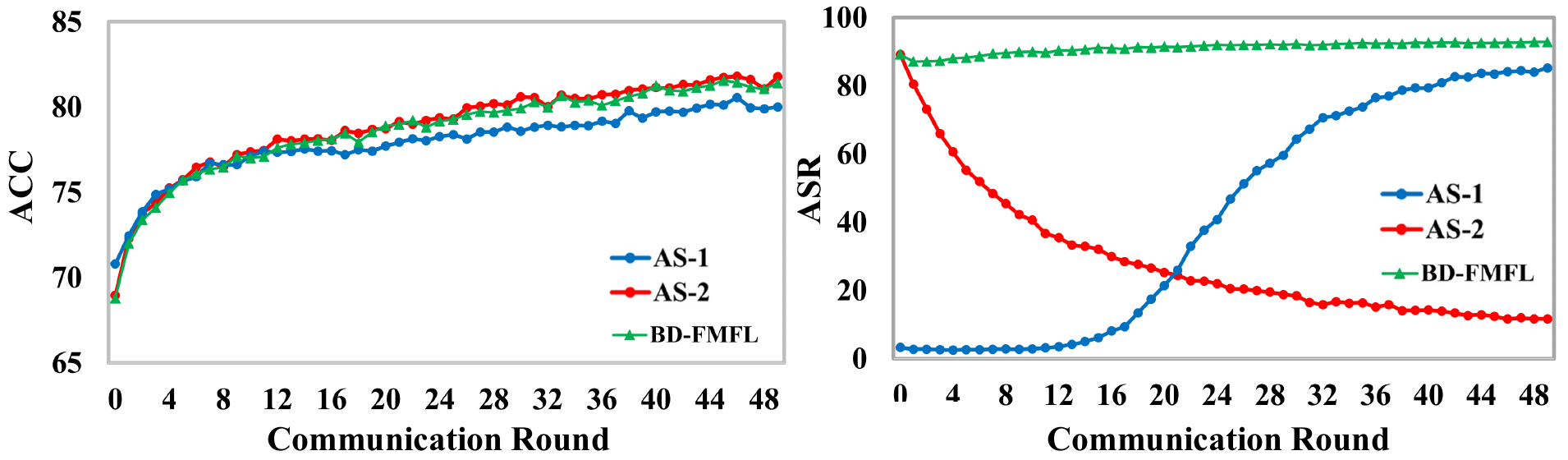}
        \caption{homo-FL scenario}
        \label{fig:ablation-study}
    \end{subfigure}
    \hfill
    \begin{subfigure}{0.45\textwidth}
        \centering
        \includegraphics[width=\linewidth]{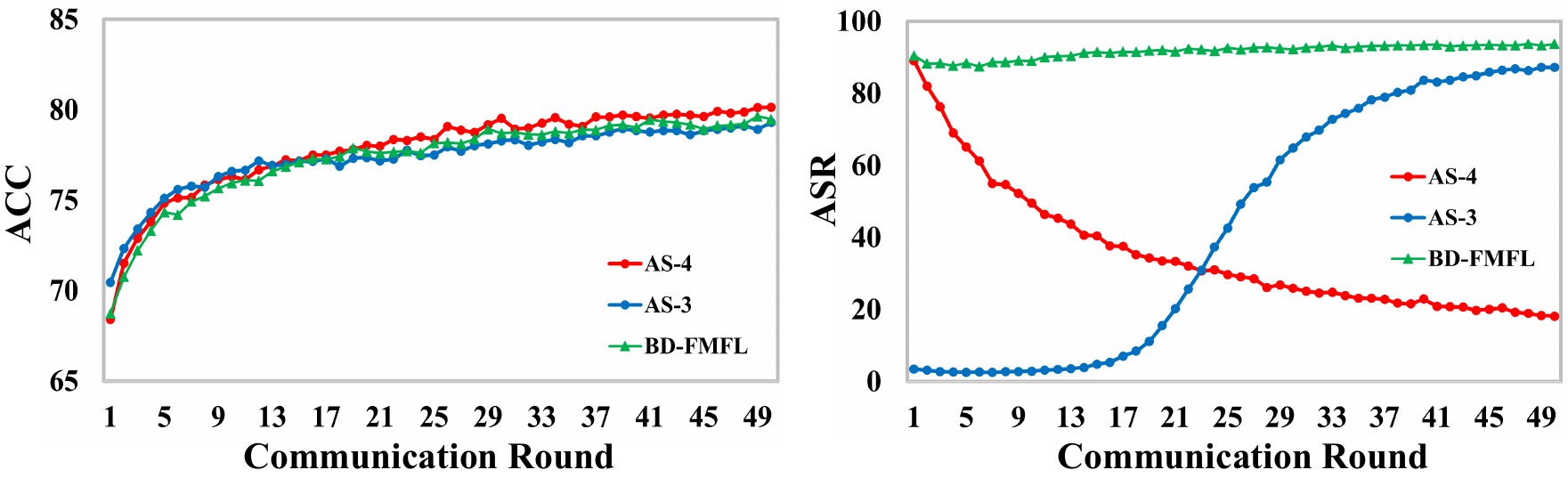}
        \caption{hete-FL scenario}
        \label{fig:ablation-study-hete}
    \end{subfigure}
    \caption{Ablation study in cross-silo FL using the IID CIFAR-10. AS-1/3: Utilizes poisoned synthetic data exclusively in ensemble distillation. AS-2/4:  Utilizes poisoned synthetic data exclusively in model initialization.}
    \vspace{-0.2in}
\end{figure}

\subsection{Ablation Study}\label{sec:AS}
BD-FMFL leverages poisoned synthetic data during both model initialization and iterative knowledge distillation. We perform an ablation study (Fig.~\ref{fig:ablation-study}) in a cross-silo homo-FL and hete-FL settings with the IID CIFAR-10 dataset to evaluate the impact of compromising each stage separately.

\textbf{AS-1: Threat Planting in Initialization.}  
To assess the role of threat planting during initialization, we introduce $\text{BD-FMFL}_\text{no-init}$, where the poisoned synthetic dataset is only used for ensemble distillation, and a clean version synthetic dataset (without trigger instances) is used for initialization. 
Fig.~\ref{fig:ablation-study} shows both attacks minimally affect ACC. BD-FMFL maintains an ASR above 80\% throughout training, while $\text{BD-FMFL}_\text{no-init}$ takes 40 rounds to achieve the same ASR, as uncorrupted initial models struggle to align with triggered instances. Over time, the contaminated synthetic data gradually corrupts the client models.

\textbf{AS-2: Threat Reinforcement via Mutual Distillation.}  
We evaluate the effect of iterative malicious knowledge distillation by introducing $\text{BD-FMFL}_\text{no-KD}$, where poisoned synthetic datasets are used only in initialization, with clean data for ensemble distillation. 
As seen in Fig.~\ref{fig:ablation-study}, both attack have similar influence on ACC. 
$\text{BD-FMFL}_\text{no-KD}$ demonstrate efficacy in the initial stages but its ASR gradually declines to 10\% as training progresses. 
The absence of iterative reinforcement weakens the attack's impact, as local fine-tuning on clean data mitigates the threat, leading to eventual forgetting by convergence. Similar results are observed in hete-FL settings (Fig.~\ref{fig:ablation-study-hete}).

\vspace{-1mm}
\subsection{Hyper-parameter Study}\label{sec:hyper-para}
\begin{figure*}
    \centering
    \includegraphics[width=1\linewidth]{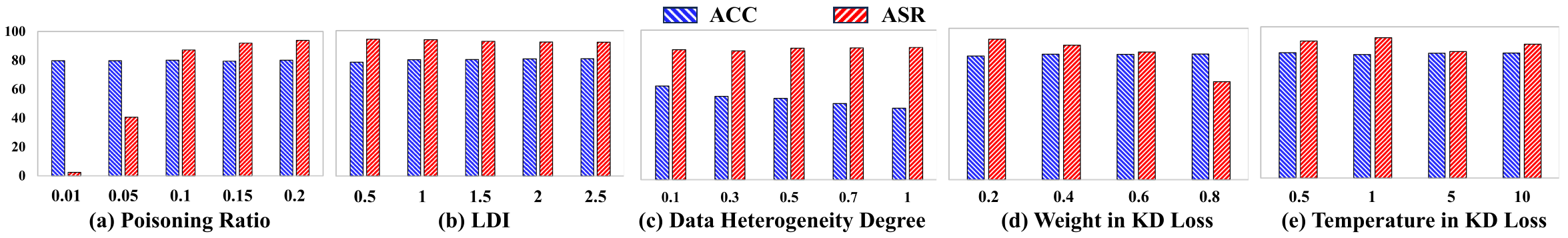}
    \caption{Hyper-parameter study in cross-silo homo-FL scenarios. (a)(b) use the IID CIFAR-10 dataset, (c) uses the non-IID CIFAR-100 dataset, (d)(e) use the IID CIFAR-10 dataset. LDI refers to the ratio between the number of iterations of (client) local training and that of (server) knowledge distillation.}
    \label{fig:hyper-para}
    \vspace{-5mm}
\end{figure*}

We analyze five key factors influencing BD-FMFL's impact on FM-FL vulnerability in cross-silo homo-FL settings. 
The results suggest that \textit{the effectiveness of the novel threat is not sensitive to the hyper-parameter settings of FL}, highlighting the importance of advanced robust FM-FL systems.

\textbf{Poisoning Rate.}  
We vary the poisoning rate of the synthetic data at 0.01, 0.05, 0.1, 0.15, and 0.2. 
Fig.~\ref{fig:hyper-para}(a) shows that when the poisoning rate exceeds 0.1, the attack becomes effective with an ASR exceeding 80\%, while ACC remains largely unaffected.

\textbf{Local-Distillation Iteration (LDI) Ratio.}  
The LDI ratio, defined as the ratio of client-side local training epochs to server-side distillation epochs per communication round, is tested at values 0.5, 1, 1.5, 2, and 2.5 (default = 1). 
As shown in Fig. \ref{fig:hyper-para} (b), the ASR decreases slightly as the LDI ratio increases, yet remains above 80\%.

\textbf{Data Heterogeneity Degree.}  
The impact of non-IID data distribution is studied using Dirichlet parameter $\beta = 0.1, 0.3, 0.5, 0.7, 1$. 
Fig.~\ref{fig:hyper-para}(c) shows that ACC decreases with increasing $\beta$, as more balanced local data distribution makes learning scarce classes harder.
The ASR remains high under different settings. 

\textbf{Weight Factor in KD Loss.}  
We test the weight factor $\alpha$ in Eq.~\ref{eq:ensemble distillation} at 0.2, 0.4, 0.6, and 0.8 (default = 0.2). 
Fig.~\ref{fig:hyper-para}(d) shows ACC is largely unaffected by $\alpha$, while ASR drops from 85\% to 65\% when $\alpha$ increases beyond 0.6. The attack remains highly effective at typical $\alpha$ values.

\textbf{Temperature in KD Loss.}  
Temperature $\tau$ (Eq.~\ref{eq:ensemble distillation}) regulates the softness of teacher model logits. Evaluated at 0.5, 1, 5, and 10 (default = 1), Fig.~\ref{fig:hyper-para}(e) shows both ACC and ASR remain largely unaffected by $\tau$ variations.

\vspace{-0.05in}
\subsection{Performance Evaluation on Text Dataset}
As shown in Tab.~\ref{tab:ours_text_attack_defense}, we evaluate the vulnerability of FM-FL systems and robustness of the existing FL backdoor defenses under the proposed attack strategy on text classification.
Here we consider both homogeneous and heterogeneous FL systems in the cross-silo setting using both IID and non-IID AG-NEWS datasets.
The results are consistent with those in the image classification task. 
The vanilla FM-FL is highly vulnerable to BD-FMFL, with ASR higher than 70\%.
Moreover, all the defense methods exhibit insufficient robustness against the proposed attack approach.
The average ASR drops less than 5\% when using \textbf{NormThr} and less than 3\% when using \textbf{Krum}. 
Using \textbf{DP}, the ASR decreases by about 30\%, and the average ACC also falls by over 10\% due to Gaussian noise introduced into the global model. This defense method experiences a significant reduction in ACC, especially in heterogeneous FL scenarios. The \textbf{Pruning} defense method remains the most effective among all defense mechanisms. The average ASR has been controlled to around 60\%. 

\begin{table}[ht]
\scriptsize
    \centering
    \caption{Vulnerability of FM-FL systems and robustness of current FL defenses against the novel attack strategy in cross-silo scenarios using the AG-NEWS dataset.}
    \vspace{-1ex}
    \begin{tabular}{p{.85cm}|p{.25cm}p{.36cm}p{.25cm}p{.36cm}p{.25cm}p{.36cm}p{.25cm}p{.36cm}p{.25cm}p{.36cm}}
    \toprule
        \textbf{Setting} & \multicolumn{2}{c}{\textbf{Vanilla}} & \multicolumn{2}{c}{\textbf{NormThr}} & \multicolumn{2}{c}{\textbf{DP}} & 
        \multicolumn{2}{c}{\textbf{Krum}} & \multicolumn{2}{c}{\textbf{Pruning}} \\
        & ACC & ASR & ACC$\downarrow$ & ASR & ACC$\downarrow$ & ASR & ACC$\downarrow$ & ASR & ACC$\downarrow$ & ASR \\
        \hline
        \multicolumn{11}{c}{\textbf{Homo-FL}} \\ 
        \hline
        \textbf{IID} & 89.73 & 76.07 & 2.13 & 71.34 & 11.50 & 40.25 & 1.11 & 75.21 & 0.31 & 37.81 \\
        \textbf{non-IID} & 96.26 & 71.00 & 0.78 & 66.83 & 8.97 & 38.76 & 0.45 & 69.87 & 1.06 & 65.66 \\
        \hline
        \multicolumn{11}{c}{\textbf{Hete-FL}} \\
        \hline
        \textbf{IID} & 89.03 & 79.17 & 0.92 & 78.56 & 16.57 & 43.94 & 0.48 & 76.27 & 2.05 & 62.88 \\
        \textbf{non-IID} & 95.75 & 76.96 & 1.41 & 74.60 & 14.51 & 50.83 & 7.31 & 64.29 & 0.87 & 71.17 \\
    \bottomrule
    \end{tabular}
    \label{tab:ours_text_attack_defense}
    \vspace{-0.05in}
\end{table}
\section{Conclusion}
In this paper, we propose a novel attack strategy that utilizes the inherent security issues to compromise the FL client models.
We specialize the strategy to backdoor attacks and conduct the first comprehensive evaluation of the vulnerability of the FM-FL under novel threats.
Our study, employing a range of established models and benchmark datasets in both image and text domains, demonstrates the significant susceptibility of FM-FL under the novel threat.
Besides, existing FL defenses offer limited protection against such threats.
Our work closes the gap in the literature investigating the robustness of FM-FL and highlights the critical need for enhanced security protocols to protect FL systems in the era of FMs.

{\small
\bibliographystyle{IEEEtran}
\bibliography{ref}}

\end{document}